\input psfig.tex  

\documentstyle[aps]{revtex}
\begin{document}

\title{Dynamic renormalization-group approach to diffusive flow in
heterogeneous systems}

\date{\today}
\author{Y\"uksel G\"unal and P B Visscher 
\footnote
{email gunal@pi.ph.ua.edu, pvissche@ph.as.ua.edu}
}
\address{Dept. of Physics, University of Alabama, Tuscaloosa, AL 
35487-0324} 

\maketitle

\begin{abstract}
Conventional methods for the simulation of diffusive systems are
quite slow when applied to strongly inhomogeneous systems. 
We present a new hierarchical approach based on dynamic 
renormalization-group ideas and on the Walsh transform (or Haar 
wavelet) of signal-processing theory.  The method is very
efficient for simulation of petroleum reservoirs or other 
strongly inhomogeneous diffusive or pressure-driven flow systems.
In a test case, the hierarchical method is found to achieve 1.5\% 
accuracy roughly 25 times faster than conventional finite 
difference methods.
\end{abstract}


Traditional finite-element and finite-difference methods for 
numerical 
solution of differential equations have a discretization error 
that depends on a power
of the time or space increment, $\Delta t$ or $\Delta r$. In the case 
of a
diffusive system, stability usually requires $\Delta t$ to be of 
order $%
\Delta r^2/D$ (where $D$ is the diffusivity), and the discretization 
error is
proportional to a power of $\Delta r$. 
In highly inhomogeneous systems, parts of which require very small 
$\Delta r$
and/or very large diffusivity $D$, this requires a very small 
$\Delta t$.

The instability for large $\Delta t$ arises when material 
moves more than one cell diameter during the time interval $\Delta t$;
in the usual explicit finite difference (EFD) algorithm, 
material is allowed to move from a cell only into its nearest 
neighbors.
Our approach solves this problem by allowing material to move across 
as many 
cells as necessary.  We describe the motion of the material by a 
discrete
Green function or influence function $G_{\Delta t}(d,s)$ such that 
$G_{\Delta t}(d,s) c(s)$ is the amount of material that moves from a 
source 
cell $s$ (whose original material content is $c(s)$) to a destination 
cell 
$d$ during the time interval $\Delta t$.  If $G_{\Delta t}(d,s)$ is 
nonzero 
only when $d$ and $s$ are nearest neighbor cells, this is equivalent 
to a 
conventional EFD algorithm.  This is the case for 
sufficiently small $\Delta t$, so we may begin with such an algorithm 
and 
coarsen the time scale by doubling $\Delta t$.  The Green function 
for the interval 
$2 \Delta t$ is obtained from that for $\Delta t$ by self-convolution 
(see 
Eq. \ref{conv} below).

Of course, this repeated convolution process increases the spatial 
range of 
the influence function, which is stored in our computer 
implementation as a 
linked list; soon the number of destination cells $d$ that can be 
reached 
from each source cell $s$ becomes large and the method becomes 
very time-consuming.  However, we can 
coarsen the space as well as the time scale, by lumping cells 
together 
into larger cells.  This decreases the number of source cells, as 
well as 
the number of destination cells reached from each source cell, and 
hence 
the size of the Green-function list.  This scheme is motivated by the 
renormalization-group method which has been so useful in the theory 
of 
critical phenomena \cite{wilson}, although our present description 
does not require prior knowledge of renormalization-group theory.  
It has been shown \cite{fixed} that the diffusion problem in a {\em 
homogeneous} system has a fixed point with respect to a combined 
space-and-time renormalization transformation.  
That is, we can continue coarsening 
the space and time scales indefinitely without indefinitely 
increasing the 
size of the Green-function list.  

Such cell coarsening is even more useful in an inhomogeneous system,
because we can use physical information to choose which cells to lump 
together.  An inhomogeneous system such as an oil reservoir tends to 
be 
compartmentalized into compartments within which oil flows fairly 
freely, 
separated by relatively impermeable regions.  If we lump cells 
between 
which there is relatively free flow (high effective diffusivity), 
when we
reach a large scale the cells will {\em be} the compartments.

\begin{figure}
\psfig{figure=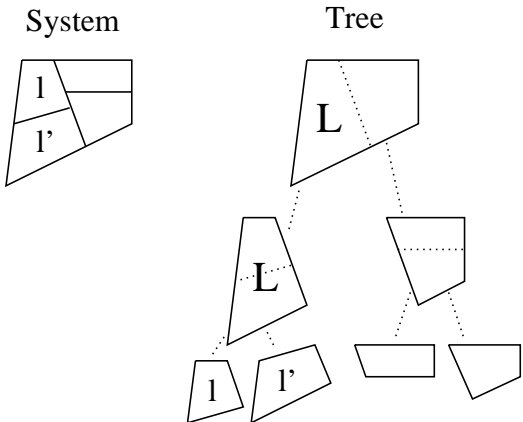}
\caption{Sketch of a hierarchically subdivided system (left) and its 
representation as a binary tree (right).  Cell $L$ is subdivided into 
$l$ and 
$l^{\prime}$ as described in the text.
} 
\label{tree}
\end{figure}

We can visualize the hierarchical lumping of cells by placing the 
cells on 
a binary tree as in Fig. \ref{tree}.

The disadvantage of this coarsening process, of course, is that we 
lose 
spatial resolution in our description of the system.  To some extent 
such a 
contraction of the description is desired, but we would like to 
maintain control of our approximations and limit the loss of 
information.
When we replace contents $c(l)$ and $c(l^{\prime})$, of two cells $l$ 
and $l'$ by 
the coarse-grained content $c(L) \equiv c(l)+c(l^{\prime})$ where the 
larger 
cell $L$ is the union of $l$ and $l^{\prime}$, we can avoid losing 
any 
information if we also include as a variable the 
difference $c(L,1) \equiv c(l)-c(l^{\prime})$ as well as 
the sum $c(L,0)$ (the arguments $0$ and $1$ simply indicate 
whether a sum or a difference is intended).

We can do this at any level of the tree shown in Fig. \ref{tree} --
the difference between the two halves of the small cell $l$ can be 
denoted
by $c(l,1)$.  We can even define a difference of differences 
$c(L,1,1) \equiv c(l,1)-c(l^{\prime},1)$.  Each "$1$" in this 
expression 
can be regarded as one bit of a "Walsh sequency index" $w$.
(The Walsh transform is a discrete signal transform used in 
electrical 
engineering\cite{wintro}\cite{wavelet}, which we here generalize to a 
hierarchical system.)
We will lump the bits into a single binary integer (Walsh index) $w$, 
so 
$c(L,1,1)$ becomes $c(L,w)$ with $w=3$ ($11$ in binary).  We can 
define a 
general "Walsh variable" recursively by 
\begin{equation}
\label{rec}
\begin{array}{c}
c(L,w0)\equiv c(l,w)+c(l^{\prime },w) \\ 
c(L,w1)\equiv c(l,w)-c(l^{\prime },w)
\end{array}
\end{equation}
whenever $l$ and $l^{\prime }$ are the two halves of the cell $L$. 
Here the Walsh index $w0$ is $w$ with a zero bit appended at the 
right, i.e., $w0\equiv 2w$; similarly $w1 \equiv 2w+1$.
In this notation, our original $c(l)$ is written $c(l,0)$ and 
serves to start the recursion.
The Walsh index plays a role similar to that of the wavenumber in the 
Fourier transform, in that variables with a small Walsh index 
describe 
large-scale structure, whereas large Walsh indices describe small 
wavelength structure within a cell $L$.
For each "$1$" bit in the binary representation of $w$, there is 
one subtraction in the construction of $c(l,w)$.

If each cell-lumping replaces two cell contents by two Walsh 
variables, 
we never lose variables and our equations remain as complicated as 
ever
(but exact).  
However, we are now in a position to selectively drop small terms in 
the 
Green function: the variables with large Walsh indices (i.e., many 
differences rather than sums) will be less important than those with 
small 
Walsh indices.  Our algorithm drops terms from the Green function 
if they are less than some preset tolerance $\delta$.  
Typically, terms involving differences are much smaller than those 
involving sums, so these are the ones dropped.  This is the virtue of 
the 
hierarchical description: terms describing the effects of a cell 
content 
are never negligible compared to the terms for nearby cells, whereas 
terms 
describing the effects of differences may be negligible compared to 
terms 
for sums.  These advantages are similar to those of spectral (Fourier 
transform) methods in homogeneous systems; in a sense one can regard 
the 
Walsh transform as the proper generalization of the Fourier transform 
to 
inhomogeneous systems.

Commercial reservoir simulation programs usually treat a reservoir as 
a
three-component system\cite{simbook} (oil, gas, water) 
in which flow is governed by
Darcy's law. However, to provide a straightforward test of the 
hierarchical
method described above, we will consider only one component (oil). In 
this
case, Darcy's law reduces to the diffusion equation
\begin{equation}
\frac{dP({\bf r},t)}{dt}=\nabla \cdot [D({\bf r})\nabla P({\bf r},t)]
\label{diffe} 
\end{equation}
where the pressure $P$ plays the role of the diffusing density;  the
effective diffusivity $D({\bf r})$ is proportional to the 
permeability. So
the problem we will actually solve is that of diffusion in a very
inhomogeneous system; the inhomogeneity is contained in the function 
$D({\bf r})$.

To describe the evolution of the cell content $c_t(d)$ of a cell 
labeled $d$
(proportional to the
"density" $P$), the most straightforward discretization of Eq. 
\ref{diffe} is
\begin{equation}
\label{disc}
\frac{c_{t+\Delta t}(d)-c_t(d)}{\Delta t}=\Delta
r^{-2}\sum_fD(f)[c_t(d_{+}(f))-c_t(d)] 
\end{equation}
where the sum is over faces $f$ of the cell $d$, and $d_{+}(f)$ is 
the cell
in front of the (directed) face $f$ (the cell behind it is always 
$d$). When
we lump cells, so some of our variables are Walsh variables $c(d,w)$, 
we
can write the discretization (Eq. \ref{disc}) in the form  
\begin{equation}
\label{ww}
c_{t+\Delta t}(d,w)=\sum_{s,v}G_{\Delta t}(d,w;s,v)c_t(s,v)\label{gd} 
\end{equation}
where $G_{\Delta t}(d,w;s,v)$ is a Green function or influence 
function describing the
influence of a Walsh variable in the source cell $s$ on one in the
destination cell $d$. Before any lumping has occurred, all Walsh 
variables
have $w=0$, and $G_{\Delta t}(d,w;s,v)=(\Delta t/\Delta r^2)D(f)$ if 
$d$ and $s$ are
neighbors separated by the face $f$ and $w=v=0$, and zero otherwise.
Equation \ref{ww} is thus equivalent to an EFD algorithm, which 
requires a small $\Delta t$; we have used the practical limit of 
stability\cite{fd},
\begin{equation}
\label{dt}
\Delta t = \Delta r^2/4D_{\text{max}}
\end{equation}
where $D_{\text{max}}$ is the maximum diffusivity in the system.

\begin{figure}[htpb]
\psfig{figure=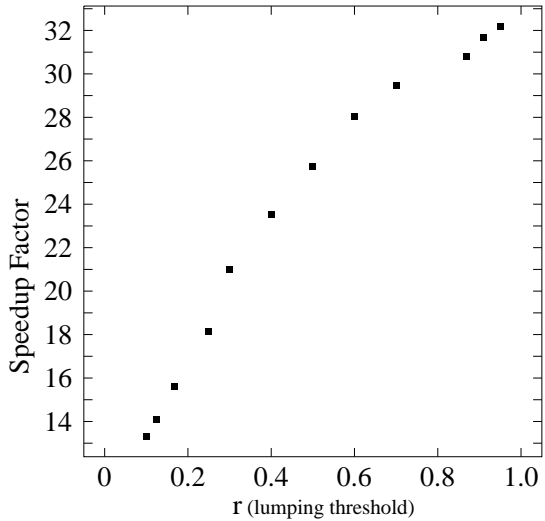,rheight=252pt,rwidth=252pt}
\caption{Speedup factor as a function of the lumping threshold 
$r$ (Eq. \protect{\ref{r}}), 
in a $32 \times 32$ system with inhomogeneity $I=20$
and tolerance $\delta = 10^{-4}$.} 
\label{rdep}
\end{figure}

We now increase $\Delta t$ to $2\Delta t$; the new Green function 
is the convolution
\begin{equation}
\label{conv}
G_{2\Delta t}(d,w;s,v)= 
\sum_{e,u} G_{\Delta t}(d,w;e,u) G_{\Delta t}(e,u;s,v)
\end{equation}
After several such convolutions, the spatial range of the Green 
function is increased, especially in regions of high diffusivity.  
Here the pressure (i.e., density) in nearby cells equalizes 
rapidly -- the contents $c(l)$ and $c(l')$ of two nearby cells 
contribute nearly equally to future contents $c(d)$: $G(d,0;l,0) 
\approx G(d,0;l',0)$ (the zeroes here are the Walsh indices).  
To decide whether two cells $l$ and $l'$ should 
be lumped together, we look at the ratio (using $d=l$)
\begin{equation}
\label{r}
r=\frac{G(l,0;l',0)}{G(l,0;l,0)}
\end{equation}
Generally $r<1$; we have used $r>0.91$ as our lumping criterion 
(see Fig. \ref{rdep}).

When we decide to lump two cells $l$ and $l'$ into a large cell 
$L$ (as in Fig. \ref{tree}), we can calculate the new Green 
function in two stages.  In the first stage we calculate elements 
$G'(d,w;s,v)$ in which the {\em destination} cell $d$ takes 
coarse values (including $L$) but $s$ takes values including $l$ 
and $l'$.  These are the same as the old 
$G$s unless $d$ is $L$, in which case we obtain from Eq. \ref{rec}
\begin{equation}
\label{lumpd}
\begin{array}{c}
G'(L,w0;s,v)=G(l,w;s,v)+G(l',w;s,v)\\
G'(L,w1;s,v)=G(l,w;s,v)-G(l',w;s,v)
\end{array}
\end{equation}
where as before $w0$ means $w$ with a zero bit appended.
In the second stage, we calculate $G''(d,w;s,v)$ where both $d$ and 
$s$ take values $L$ and not $l$ or $l'$: we coarsen the source cell.
Again, $G''(d,w;s,v)=G'(d,w;s,v)$ unless $s=L$, in which case 
\begin{equation}
\label{lumps}
\begin{array}{c}
G''(d,w;L,v0)=\frac12 [G'(d,w;s,v)+G'(d,w;s',v)]\\[0.02 in]
G''(d,w;L,v1)=\frac12 [G'(d,w;s,v)-G'(d,w;s',v)]
\end{array}
\end{equation}

Although we have implemented our algorithm in three
dimensions, we use a 2D system for our test calculation
The test system has four control parameters: $N$, 
$I$, $\delta$, and $r$.  The first two describe the complexity of 
the system: $N$ is the system size ($16 \times 16$, $32 \times 
32$, or $64 \times 64$) and the inhomogeneity parameter $I$ is 
a normalized standard deviation: 
the standard deviation of the permeability divided by the mean 
permeability. 
The other two parameters are the error tolerance $\delta$, 
which we choose to give an acceptable overall truncation error,  
and the lumping threshold $r$ (Eq. \ref{r}), 
which we tune to maximize the speed of the algorithm.

\begin{figure}
\psfig{figure=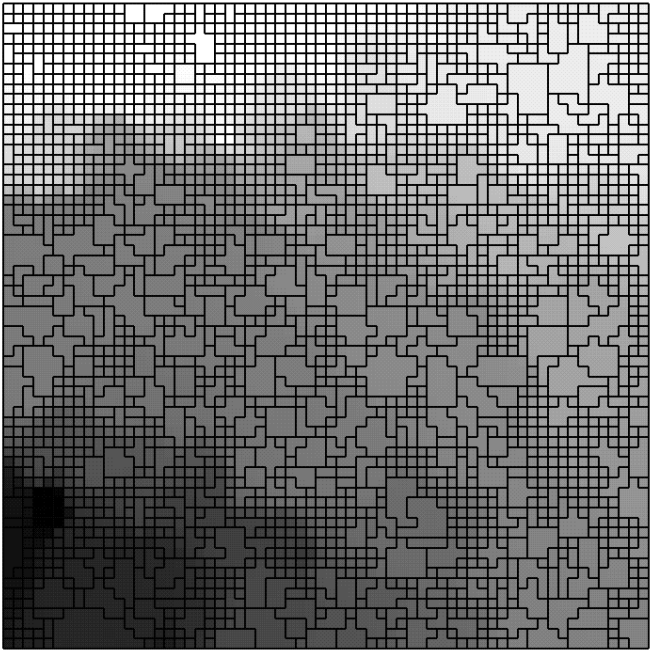}
\caption{The final density distribution in the $64 \times 64$ 
system with inhomogeneity $I=20$, showing boundaries of lumped cells.
} 
\label{final}
\end{figure}

The diffusivity (i.e., permeability) distribution we have used is
a realization of a log-normal distribution
with fractal spatial correlations, obtained by exponentiating a 
correlated Gaussian distribution described elsewhere\cite{dye}. 
We adjust the normalized standard deviation $I$ by scaling 
the Gaussian distribution before exponentiating it.  The 
prefactor that governs the overall scale of the diffusivity or 
permeability can be removed from the problem by rescaling time.
The system shown in Fig. \ref{final} is $64 \times 64$; 
we specify the permeability in the smaller-$N$ systems by coarse-
graining (averaging over $2 \times 2$ or $4 \times 4$ cells).
The permeability at a face is taken to be the average of that in the 
adjoining cells.

As a test initial condition, we use a delta-function density
concentrated in the lower left cell of the system.  
After an infinite time, the density takes a uniform value 
$P_\infty$; we evolve the system until the density in the source 
cell is $2P_\infty$, as shown in Fig. \ref{final}.

\begin{figure}
\psfig{figure=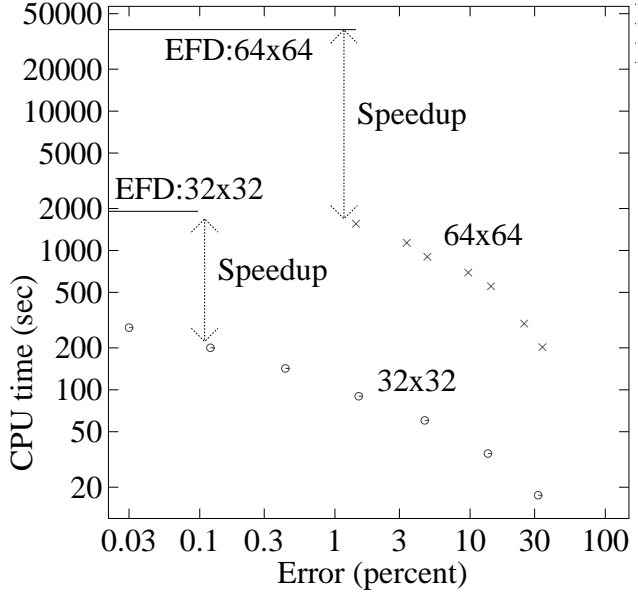,rheight=252pt,rwidth=252pt}
\caption{Logarithmic plot of the CPU time required by our 
hierarchical Green function algorithm, compared to that required by
an explicit finite 
difference (EFD) algorithm, for a $64 \times 64$ system with
inhomogeneity $I=20$.  The rightmost data point for each system 
size has tolerance parameter $\delta=10^{-2}$, and those for 
$32\times 32$ system decrease by factors of $10$.  For $64 \times 64$ 
there are points at $5\times 10^{-n}$ as well as at $10^{-n}$.
} 
\label{fd}
\end{figure}

To compare our scheme with an EFD algorithm, 
we look first at the $64 \times 64$ system 
with a substantial inhomogeneity ($I=20$), and 
use $r=0.9$ (justified by Fig. \ref{rdep} below).  We vary the 
remaining parameter, the tolerance $\delta$, and plot in Fig. 
\ref{fd} the required CPU time (on a Silicon Graphics R4000PC 
Indy, 133 MHz) against the accuracy achieved, 
defined by the fractional rms error
\begin{equation}
\label{error}
\text{error}^2 \equiv 
N^{-1} \sum_c [\frac{P(c)-P_{exact}(c)}{P_\infty}]^2.
\end{equation} 

Note that the speedup factor of our algorithm compared to the EFD 
algorithm 
increases rapidly as the allowable error is increased.  It is 
indicated by 
a double arrow at the error value of $1.5\%$, where it is about 
$25$.  Using this factor as a figure of merit for our algorithm, we 
plot it in Fig. \ref{rdep} as a function of the lumping parameter 
$r$; 
evidently performance improves as we increase $r$ toward $1.0$.  
Placing $r$ very close to $1.0$ risks magnifying the effects of small 
numerical errors, so we used $r=0.9$ in the other figures.

We are using the explicit finite difference algorithm as our 
point of comparison, not because it is the most efficient 
existing algorithm, but because it is the simplest.  However, the 
use of more efficient implicit algorithms 
increases the allowed $\Delta t$ only by factors of order $2$ 
and does not affect our conclusions.

The dependence of the speedup factor on the system properties is 
shown in Table \ref{IN}.  Evidently the Green function algorithm 
is most advantageous in highly inhomogeneous systems.
\begin{table}[h]
\begin{tabular}{c|cccc} 
& \multicolumn{3}{c}{System Size}\\ \cline{2-5}
$I$        & $8 \times 8$ &$16\times 16$&$32\times 32$&$64\times 64$ 
\\ \hline

10         & 0.2 & 1.6& 15& 18   \\ 
20         & 0.3 & 3.0& 32& 43  \\ 
40         & 0.3 & 5.2& 63& 71  \\ 
\end{tabular}
\caption{Speedup factor as a function of system size $N$, for 
three values of the inhomogeneity $I$.  Tolerance is $\delta = 
10^{-4}$.}
\label{IN}
\end{table}

Although the method described here has some features in common with 
methods 
already in common use in grid-based numerical simulation, none of 
these 
older methods approaches its efficiency for inhomogeneous systems.  
The idea 
of coarsening cells is used in "multigrid" methods \cite{multigrid}.  
For 
example, the solution of Laplace's equation by the 
relaxation method is very slow on a fine grid.  
It can be speeded up by doing a few iterations of relaxation 
on a larger grid to get the coarse features of the solution 
correct, and then returning to the fine grid to improve the finer 
features.  
Computational fluid dynamics codes often use an "adaptive grid" 
method 
\cite{adapt}, 
wherein larger grid sizes are used in regions where fields do not 
vary 
rapidly in space, and finer grids are used where there are fine-scale 
variations in the fields.  These regions are typically rectangular 
and 
cannot follow compartment shapes as closely as ours.  Unlike in our 
approach, the time increment cannot be increased above what is stable 
on the 
finest grid.  

In conclusion, we have shown that a hierarchical algorithm based on 
the 
dynamic renormalization group and the Walsh transform can simulate 
diffusive flow in an inhomogeneous system much more efficiently than 
conventional finite-difference algorithms.  This occurs because the 
rapid power-law dependence of CPU time on the fundamental scales 
$\Delta r$ and $\Delta t$ is replaced in the hierarchical method by 
a logarithmic dependence.

\vspace{0.2 in}
The work described here was partially supported by the DOE under 
Cooperative
Agreement DE-FC02-91ER75678.

\end{document}